\documentclass[preprintnumbers,amsmath,amssymb,floatfix,nofootinbib,aps]{revtex4}
\usepackage[dvips]{graphicx,graphics}
\usepackage{dcolumn}
\usepackage{bm}
\usepackage{epsfig}

\def\beq{\begin{equation}}
\def\eeq#1{\label{#1}\end{equation}}
\def\eeqn{\end{equation}}

\def\beqa{\begin{eqnarray}}
\def\eeqa#1{\label{#1}\end{eqnarray}}
\def\eeqan{\end{eqnarray}}

\let\bar=\overbar

\def\Dslash{\not{\hbox{\kern-4pt $D$}}}
\def\dslash{\not{\hbox{\kern-2pt $\del$}}}

\def\msb{{\bar{\ssstyle M \kern -1pt S}}}

\usepackage{fancyhdr,graphicx}
\newcommand{\dirac}{\partial\llap{$\diagup$\kern-2pt}}

\def\Title#1{\begin{center} {\Large {\bf #1} } \end{center}}

\begin{document}

\Title{Multiple critical point structure for chiral phase transition
induced by charge neutrality and vector interaction}

\bigskip\bigskip

%+\addcontentsline{toc}{chapter}{{\it I. Med}}
%+\label{MedInseneStart}

\begin{raggedright}

{\it Zhao Zhang\index{Zhang, Z.} and  \it Teiji Kunihiro\index{Kunihiro, T.}\\
Department of Physics\\
Kyoto University\\
Kyoto 606-8502\\
Japan\\
{\tt Email: zhaozhang@pku.org.cn}}
\bigskip\bigskip
\end{raggedright}

The combined effect of the repulsive vector interaction and the
positive electric chemical potential on the chiral phase
transition is investigated  by considering neutral color
superconductivity. Under the charge-neutrality constraint, the
chiral condensate, diquark condensate and quark number densities
are obtained in two-plus-one-flavor Nambu-Jona-Lasinio model with
the so called Kobayashi-Maskawa-'t Hooft term. We demonstrate that
multiple chiral critical-point structures always exist in the
Nambu-Jona-Lasinio model within the self-consistent mean-field
approximation, and that the number of chiral critical points can
vary from zero to four, which is dependent on the magnitudes of
vector interaction and the diquark coupling.

\section{Introduction}

It is generally believed that QCD exhibits a rich phase structure
in an extreme environment such as at high temperature and high
baryon chemical potential. For the chiral phase transition, it is
a widely accepted view that the critical point(s) (i.e.\ the end
point of the first-order phase boundary) should exist at finite
temperature and density. Usually, a schematic $T$-$\mu$ phase
diagram with one critical point is widely adopted in the
literature~\cite{Stephanov:2007fk}.

%It is extensively believed that
The color-superconducting (CSC)
phase may appear in the $T$-$\mu$ phase diagram of QCD for low
temperature and large baryon chemical
potential~\cite{WilczekReview,RischkeReview,BuballaReview,AlfordReview}.
At asymptotically high density that justifies the perturbative QCD
calculations, the color-flavor locked (CFL) phase~\cite{CFL} has
been established as the ground state of quark matter. For the
intermediate density region which may exist in the core of a
compact star, the nonperturbative features of QCD play a more
important role in the phase structure of QCD, and both CFL and
non-CFL CSC phases may manifest themselves in this region.

The appearance of the CSC phase around the boundary of the chiral
phase transition at low temperatures should affect the chiral
phase transition,  which
%and hence the interplay between the
%chiral condensate and the diquark condensate
may result in an unexpected phase structure of QCD.  Especially,
the competition between the chiral condensate and the diquark
condensate may lead to the emergence of the new chiral critical
point(s) in the low temperature region. Such an example was first
presented in~\cite{KitazawaVector} in a two-flavor
Nambu-Jona-Lasinio (NJL) model with the vector interaction : It
was found that the repulsive vector interaction can lead to a
two-critical-point structure in the $T$-$\mu$ phase diagram of
QCD.  The appearance of the new critical point is attributed to
the fact that the density-density correlation induced by the
vector interaction effectively enhances the competition mentioned
above while weakening the first-order chiral restoration. For the
three flavor case, the realization of a similar QCD phase diagram
has been recently conjectured ~\cite{Hatsuda:2006ps}, where the
$\mathrm{U_A}(1)$-breaking vertex may induce a new critical point
in the low temperature region. This conjecture is based on a
general Ginzburg-Landau theory constrained by QCD symmetries and
the CFL phase is assumed to appear near the chiral boundary in the
$SU_f(3)$ limit. It has been argued that the resultant crossover
of the chiral restoration at small temperatures embodies the
hadron-quark continuity hypothesis~\cite{Continuity}.

Note that the two-critical point structures obtained above are all
based on the  exact flavor symmetry. Namely, the
quarks with different flavor have the same mass and density. In
 realistic situations, $s$ quark is much heavier than $u$ and $d$ quarks,
which may make a different phase structure involving the chiral
transition. In particular, the large mass difference between $s$
quark and $u(d)$ quark may disfavor the formation of the CFL in
the intermediate density region. Thus, the conjecture of the
two-critical-point structure claimed in ~\cite{Hatsuda:2006ps} may
not be true in the realistic situation. In addition,  near the
chiral boundary, even though the mass difference between the $u$
quark and the $d$ quark is very small in contrast to the quark
chemical potential,  the constraint by local electric-charge
neutrality may lead to a relatively large density disparity
between them.
% Therefore, in general,
In short,  we can expect that the
Fermi sphere differences among $u$, $d$ and $s$ quarks should have
significant influence on the chiral phase transition in the
presence of CSC.

Considering the local charge neutrality constraint, it was first
disclosed in ~\cite{Zhang:2008wx} that the positive electric
chemical potential $\mu_e$ plays a similar role on the chiral
transition as the repulsive vector interaction. This is another
mechanism which can realize a multiple chiral critical-point
structure.  In a simple two-flavor NJL model, the same
two-critical-point structure found in ~\cite{KitazawaVector} can
be realized without introducing vector interaction when taking
into account the local electric-neutrality~\cite{Zhang:2008wx}.
Moreover, besides its role as an effective vector interaction, the
positive $\mu_e$ also implies a finite disparity between Fermi
spheres of $u$ and $d$ quarks. For a system composed of two
different Fermions with the mismatched Fermi spheres, it is known
that the energy gap of the Cooper pairing between them can {\em
increase} with temperature. This unusual thermal behavior is due
to the smearing of the Fermi surface by the temperature. Such an
unconventional behavior in the two-flavor neutral CSC phase can
lead to a special competition between the chiral condensate and
the diquark condensate, which may be enhanced with increasing
temperature. For some model parameter regions, this abnormal
competition can give rise to {\em three} chiral critical structure
in the phase diagram~\cite{Zhang:2008wx}. We remark that the
interplay of two order parameters with the external constraint(s)
should be a general mechanism for a realization of a multiple
critical-point structure in many-body system including those in
condensed matter physics.

It is an interesting problem whether the multiple critical-point
structure can appear in the QCD phase diagram for the
two-plus-one-flavor case. The CFL phase may be disfavored near the
chiral boundary in this case due to the relatively heavy strange
quark. In addition, the $\mathrm{U_A}(1)$ anomaly of QCD may have
significant influence on the chiral phase transition. Such an
anomaly can be successfully fulfilled in the effective quark field
theory with the so-called Kobayashi-Maskawa-'t Hooft(KMT)
term\cite{KMT}. Because of the similar roles of the repulsive
vector interaction and the electric chemical potential under the
neutrality constraint on the chiral phase transition, it is
naturally expected that chiral restoration is weakened more
significantly when taking into account both aspects
simultaneously. The study of the QCD phase diagram including all
these ingredients has been done in \cite{Zhang:2009zk} in the
framework of a two-plus-one-flavor NJL model.

As we will show below, the multiple chiral critical point
structures can still emerge in the QCD phase diagram for the
two-plus-one-flavor case. Such a conclusion is qualitatively
different from the result in \cite{Ruester:2005jc,Abuki:2005ms}
where the same model was used without considering the vector
interaction. This just reflects the combined effect of the
electric chemical potential and vector interaction on the chiral
phase transition through a special competition of the chiral
condensate and the diquark condensate. It should be stressed that
the present mechanism giving rise to the multiple chiral critical
structures is different from that reported in
\cite{Hatsuda:2006ps} where the anomaly term in the flavor-$SU(3)$
limit plays the essential role. In our case,  only two-flavor CSC
(2CSC) phase appears near the chiral boundary and there is no
cubic interactions between the chiral condensate and the diquark
condensate and the number of the chiral critical points can be
more than two.

\section{Two-plus-one-flavor NJL Model}

In this part, for simplicity, a local two-plus-one-flavor NJL
model is adopted to investigate the QCD critical-point structure.
The two-plus-one-flavor NJL model was developed in the mid-1980s
\cite{3-NJL1,3-NJL2,3-NJL3}, and the most popular version includes
a chiral symmetric four-quark interaction term and a determinantal
term \cite{KMT} in flavor space
\cite{3-NJL4,3-NJL5,Rehberg:1995kh}. To compare with the previous
study, we take the same model parameters as in
Ref.\cite{Ruester:2005jc} by including the vector interaction
channel. The corresponding Lagrangian density is given by

\begin{eqnarray}
\mathcal{L} &=& \bar \psi \, ( i \dirac - \hat{m} \, ) \psi +G_S
\sum_{i=0}^8 \left[ \left( \bar \psi \lambda_i \psi \right)^2 +
\left( \bar \psi i \gamma_5 \lambda_i \psi \right)^2 \right]+G_D
\sum_{\gamma,c} \left[\bar{\psi}_{\alpha}^{a} i \gamma_5
\epsilon^{\alpha \beta \gamma} \epsilon_{abc} (\psi_C)_{\beta}^{b}
\right] \left[ (\bar{\psi}_C)_{\rho}^{r} i \gamma_5 \epsilon^{\rho
\sigma \gamma} \epsilon_{rsc} \psi_{\sigma}^{s} \right]
\nonumber \\
&-& G_V\sum_{i=0}^8 \left[ \left( \bar \psi \gamma^\mu \lambda_i
\psi \right)^2 + \left( \bar \psi i \gamma^\mu \gamma_5 \lambda_i
\psi \right)^2 \right]-K \left\{ \det_{f}\left[ \bar \psi \left( 1
+ \gamma_5 \right) \psi \right] + \det_{f}\left[ \bar \psi \left(
1 - \gamma_5 \right) \psi \right] \right\}\;, \label{Lagrangian2}
\end{eqnarray}
where the quark spinor field $\psi_{\alpha}^{a}$ carries color
($a=r,g,b$) and flavor ($\alpha=u,d,s$) indices. In contrast to
the two-flavor case, the matrix of the quark current masses is
given by $\hat{m} =\text{diag}_{f}(m_u,m_d,m_s)$ and the Pauli
matrices in flavor space are replaced by the Gell-Mann matrices
$\lambda_i$ in flavor space with $i=1,\ldots,8$, and
$\lambda_0\equiv \sqrt{2/3} \,\openone_{f}$. The corresponding
parametrization of the model parameters is given in Table
\ref{tab2}, where $G_S$, the coupling constant for the scalar
meson channel, and $K$, the coupling constant responsible for the
$U_A(1)$ breaking, or the KMT term\cite{KMT}, are fixed by the
vacuum physical observables.  Because there are no reliable
constraints on the coupling constants $G_V$ and $G_D$ within the
NJL model, these two model parameters are taken as free parameters
in our treatment. The standard ratios of $G_V/G_S$ and $G_D/G_S$
from Fierz transformation are 0.5 and 0.75, respectively. In
addition, in the molecular instanton liquid model, the ratio
$G_V/G_S$ is 0.25. Considering these two points, the reasonable
value of $G_V/G_S$ may be located in the range from 0 to 0.5.

\begin{table}{
\begin{tabular}{c|c|c|c|c|c}
\hline\hline {\quad $m_{u,d}$(MeV) \quad  } &  \quad {\quad
$m_{s}$(MeV) \quad}&  \quad {\quad $G_S\Lambda^2$ \quad}& \quad
\small {\quad $K\Lambda^5$ \quad}&\quad {\quad $\Lambda$ (MeV)}
\quad &{$M_{u,d}$ (MeV)}\\
\hline 5.5 & 140.7 & 1.835 & 12.36 & 602.3 & 367.7 \\
\hline\hline {\quad $f_{\pi}$(MeV) \quad  } & \quad {\quad
$m_{\pi}$(MeV) \quad} & \quad {\quad $m_K$ (MeV) \quad}& \quad
{\quad $m_{\eta^{,}}$(MeV) \quad} & \quad {\quad $m_{\eta}$(MeV)}
\quad& {$M_{s}$ (MeV)}\\
\hline 92.4 & 135 & 497.7 & 957.8 & 514.8  & 549.5\\
\hline\hline
\end{tabular}
\caption{Model parametrization of the two-plus-one-flavor NJL
model.} \label{tab2}}
\end{table}

\section{Thermodynamic potential for neutral color superconductivity}

In general, there exist nine possible two-quark condensates for
the two-plus-one-flavor case with the Lagrangian
(\ref{Lagrangian2}): three chiral condensates $\sigma_\alpha$,
three diquark condensates $\Delta_c$,  and three vector quark
condensates $\rho_\alpha$, where $\alpha$ and $c$
 stand for three flavors and three colors, respectively.
In the mean-field level, the thermodynamic potential for the
two-plus-one-flavor NJL model  including the charge-neutrality
constraints, is given by
\begin{eqnarray}
\Omega &=& \Omega_{l} + \frac{1}{4 G_D} \sum_{c=1}^{3} \left|
\Delta_c \right|^2-2 G_V \sum_{\alpha=1}^{3} \rho_\alpha^2
+2 G_S \sum_{\alpha=1}^{3} \sigma_\alpha^2 \nonumber\\
&-& 4 K \sigma_u \sigma_d \sigma_s -\frac{T}{2V} \sum_K \ln \det
\frac{S^{-1}_{MF}}{T} \; , \label{Omega2}
\end{eqnarray}
where $\Omega_{l}$ stands for the contribution from the free
leptons. Note that,  for consistency, $\Omega_{l}$ should include
the contributions from both electrons and muons. Since
$M_\mu>>M_e$ and $M_e\approx 0$, ignoring the contribution from
muons has little effect on the phase structure. The corresponding
$\Omega_{l}$ takes the form

\begin{equation}
 \Omega_l = -\frac{1}{12\pi^2}\left(\mu_e^4+2\pi^2T^2\mu_e^2
            +\frac{7\pi^4}{15}T^4\right) \,.
\label{omega-L}
\end{equation}

It should be stressed here that,  for simplicity, the
contributions of the cubic mixing terms among  three different
condensates, such as $\sigma\Delta^2$, $\rho\Delta^2$, and
$\sigma\rho^2$,  are neglected in Eq.~(\ref{Omega2}). These terms
arise from the KMT interaction which may or may not affect the
phase structure. In particular  it was argued in
\cite{Hatsuda:2006ps} that the cubic mixing term between chiral
 and diquark condensates  may play an important role in the
chiral phase transition in the low temperature region. Beside the
direct contribution of these cubic terms to the thermodynamic
potential, the flavor mixing terms arising from the KMT
interaction also have influence on the dispersion relationship of
quasi-quarks. For example, both the diquark condensate and the
quark number density  contribute to the dynamical quark masses.
Therefore, it is a very interesting subject to investigate the
possible effect of these cubic coupling terms on the phase
diagram.
%by using dynamic models of QCD.
Leaving an investigation of this interesting problem to our future
work, here we just simply assume that none of these mixing terms
makes a qualitative difference in the phase diagram.

Because of the large mass difference between the strange quark and
the $u,\, d$ quarks , the favored phase at low temperature and
moderate density can be most probably the 2CSC phase rather than
CFL phase as is demonstrated in the two-plus-one-flavor NJL
model\cite{Ruester:2005jc, Abuki:2005ms}. These studies suggest
that the strange quark mass is close to or even larger than $\mu$
near the chiral boundary, which means that the strange quark
density is considerably smaller than that of the $u$ and $d$
quarks. Therefore, for the two-plus-one-flavor case, the light
quarks still play the dominant role at least around the phase
boundary of the chiral transition,  and the amount of the density
difference between the $u$ and $d$ quarks under the
electric-charge-neutrality constraint is still similar to that in
the two-flavor case. Since the main purpose of our study is to
investigate the influence of the neutral CSC phase on the chiral
phase transition by taking into account the vector interaction,
only  the 2CSC phase is considered in the following.

The inverse quark propagator $S_{MF}^{-1}$ in the 2CSC phase for
the two-plus-one-flavor case formally takes the same form as that
in two-flavor case \cite{Zhang:2009zk},  with the extended
matrices $\hat{\mu}$ and $\hat{M}$ in the three-flavor space. We
refer to \cite{Zhang:2009zk} for the notations to be used below.
The constituent quark mass is given by
\begin{equation}
M_\alpha = m_\alpha - 4 G_S \sigma_\alpha + 2 K \sigma_\beta
\sigma_\gamma \; , \label{Mi}
\end{equation}
and the effective quark chemical potentials take the form
\begin{eqnarray}
\tilde{\mu}_u = \mu - 4 G_V \rho_u-\frac{2}{3}\mu_e, \label{ui}\\
\tilde{\mu}_d = \mu - 4 G_V \rho_d+\frac{1}{3}\mu_e, \label{di}\\
\tilde{\mu}_s = \mu - 4 G_V \rho_s+\frac{1}{3}\mu_e. \label{si}
\end{eqnarray}
The average chemical potential between $u$ quark and $d$ quark is
defined by
\begin{equation}
 \bar{\tilde{\mu}} = \frac{\tilde{\mu}_{rd}+\tilde{\mu}_{gu}}{2}
 = \frac{\tilde{\mu}_{ru}+\tilde{\mu}_{gd}}{2} = \mu-\frac{\mu_e}{6}
 -2G_V(\rho_u+\rho_d)+ \frac{\mu_8}{3} \,,
\label{Average}
\end{equation}
and the effective mismatch between the chemical potentials of the
u quark and the d quark takes the form
\begin{equation}
\delta\tilde{\mu}=\tfrac{1}{2}(\mu_e-4G_V(\rho_d-\rho_u)).
\label{Mismatch}
\end{equation}
The quantity $\bar{\tilde{\mu}}$ ( $\delta\tilde{\mu}$ ) still has
the same form as that in two-flavor case \cite{Zhang:2009zk}.

Note that there are three main changes induced by the nonzero
vector-type quark condensates in comparison to the case without
vector interactions \cite{Zhang:2008wx}. First, it give new
negative contributions to the thermal potential, which favors the
phase with relatively larger dynamical quark mass. This effect
becomes more significant when the quark number density is sizable.
Second, it gives rise to a negative dynamical chemical potential,
which can delay the chiral restoration towards larger chemical
potential to drive the formation of the coexistence (COE) phase
with both the $\chi\text{SB}$ and the CSC phase. In the COE
region, the competition between the two order parameters can
significantly weaken the first-order chiral phase transition.
Third, the disparity between the densities of $u$ and $d$ quarks
can effectively suppress the chemical-potential mismatch between
these two flavors, which might partially or even totally cure the
chromomagnetic instability. The detailed description on the third
point was given in \cite{Zhang:2009zk}.

If the mass difference between the $u$  and $d$ quarks is ignored,
which is actually an excellent approximation
\cite{Ruester:2005jc}), the last term in Eq. (\ref{Omega2}) has an
analytical form, and hence, the numerical calculation is greatly
simplified. Adopting the variational method, we get the eight
nonlinear coupled equations
\begin{equation}
 \frac{\partial\Omega}{\partial\sigma_u}=
 \frac{\partial\Omega}{\partial\sigma_s}=
 \frac{\partial\Omega}{\partial\Delta}=
 \frac{\partial\Omega}{\partial\rho_{u}}=
 \frac{\partial\Omega}{\partial\rho_{d}}=
\frac{\partial\Omega}{\partial\rho_{s}}=
 \frac{\partial\Omega}{\partial\mu_e}=
 \frac{\partial\Omega}{\partial{\mu_8}}=0\, .
 \label{gapeq}
\end{equation}
In fact, since $\mu_8$ is found to be tiny around the critical point
of the chiral transition \cite{Ruester:2005jc,Abuki:2005ms}, we can
set $\mu_8=0$ without affecting
the numerical results in any significant way.
% and the nonlinear equations can be reduced to seven.

\section{Numerical calculation and discussion}

%Similar to the two-flavor case,
To investigate the influence of the vector interaction on the
chiral phase transition, the coupling $G_V/G_S$ is taken as a free
parameter in the following, and the diquark coupling $G_D/G_S$ is
fixed to the standard value. Because of the contribution from the
six-quark KMT interaction, the ratio $G_D/G_S$ obtained by Fierz
transformation should be 0.95 rather than 0.75 in the case where
only the four-quark interaction is considered\cite{BuballaReview}:
In this case, the effective four-quark interaction which
determines the quark constituent mass in vacuum is
$G_{S'}=G_S-\frac{1}{2}K\sigma_s$,  and the standard value of
$G_D/G_{S'}$ should be 0.75.

For convenience, the same notations as those in
Refs.~\cite{Hatsuda:2006ps,Zhang:2008wx} are adopted to
distinguish the different regions in the $T$-$\mu$ phase diagram
of QCD: NG, CSC, COE, and NOR refer to the hadronic
(Nambu-Goldstone) phase with $\sigma\neq0$ and $\Delta=0$, the
color-superconducting phase with $\Delta\neq0$ and $\sigma=0$, the
coexisting phase with $\sigma\neq0$ and $\Delta\neq0$, and the
normal phase with $\sigma=\Delta=0$, respectively, though they
only have exact meanings in the chiral limit.

\begin{figure}
\hspace{-.0\textwidth}
\begin{minipage}[t]{.5\textwidth}
\includegraphics*[width=\textwidth]{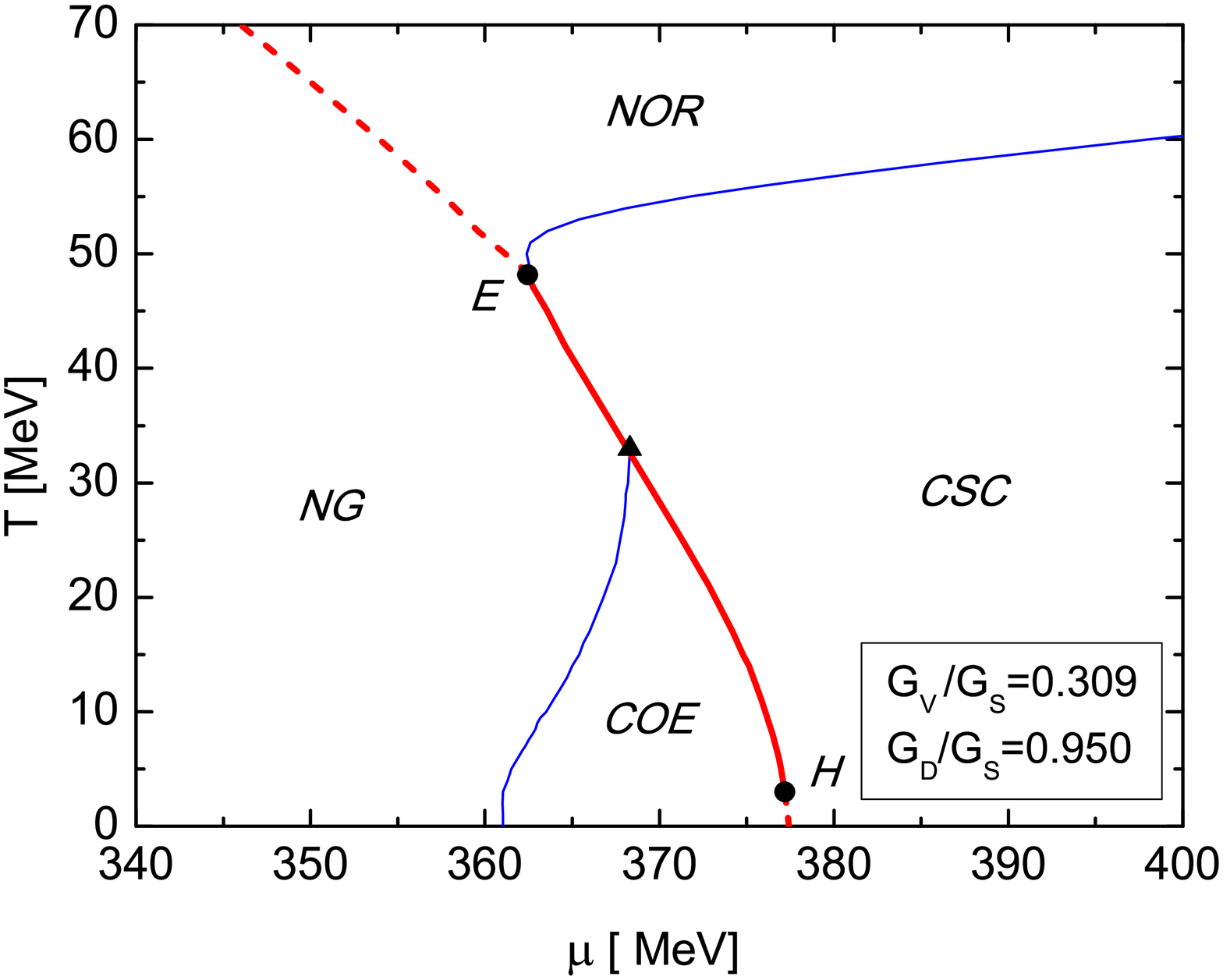}
\centerline{(a)}
\end{minipage}
\hspace{-.05\textwidth}
\begin{minipage}[t]{.5\textwidth}
\includegraphics*[width=\textwidth]{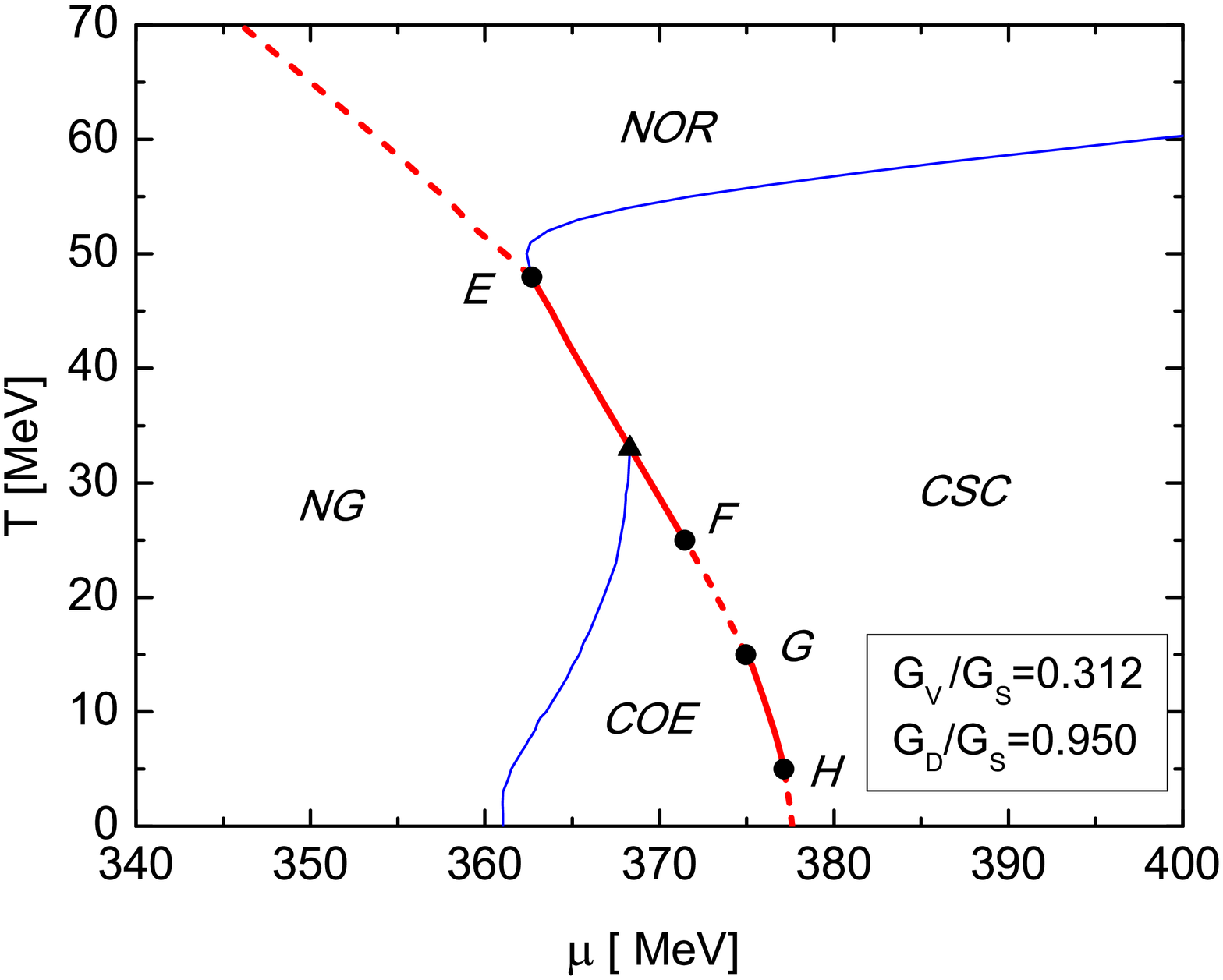}
\centerline{(b)}
\end{minipage}
\hspace{-.0\textwidth}
\begin{minipage}[t]{.5\textwidth}
\includegraphics*[width=\textwidth]{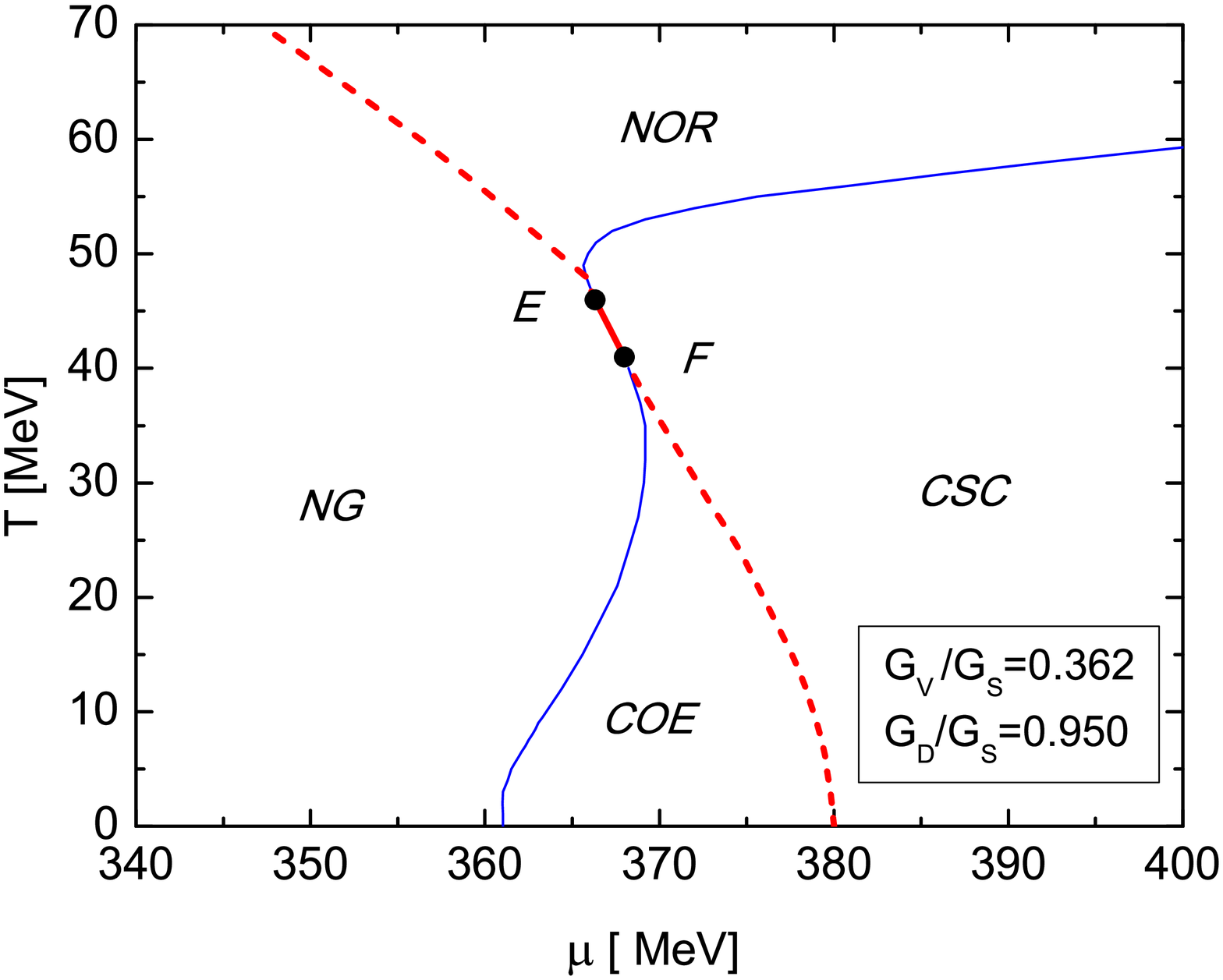}
\centerline{(c)}
\end{minipage}
\hspace{.0\textwidth} \caption{The phase diagrams for the
two-plus-one-flavor NJL model with varying $G_V/G_S$ and fixed
$G_D/G_S=0.95$. Electric-charge-neutrality is taken into account,  and
only the phase diagrams with multiple chiral critical points are
shown. } \label{fig:pd-3flavor}
\end{figure}

The phase diagrams with a multiple critical-point structure are
shown in Fig.~\ref{fig:pd-3flavor} for various strengths of the
vector interaction. As the vector interaction becomes large, the
traditional one critical-point structure is replaced by the two
critical-point structure and a new critical point at a very low
temperature emerges. Such a two critical-point structure of the
QCD phase diagram was first suggested or conjectured in
\cite{KitazawaVector} and \cite{Hatsuda:2006ps}, respectively.
With a further increase of the vector interaction, a four
critical-point structure manifest itself in the phase diagram,
which is a quite new findings. The mechanism for realizing such an
unusual structure is traced back to the nonzero $\mu_e$, which
leads to the abnormal thermal behavior of the diquark condensate
in the COE region. The phase structure with more than two critical
points was first obtained in \cite{Zhang:2008wx}. With a still
further increase of $G_V$, the number of the chiral critical
points is reduced to two from four because of the more strong
competition between the chiral condensate and diquark condensate.
When $G_V$ is large enough, the chiral transition will totaly
become crossover and there is no critical point in the phase
diagram. Fixing the ratio $G_D/G_S$ as its standard value, the
numerical calculation shows that there appear five types of
critical-point structure when the ratio of $G_V/G_S$ varing from
zero to its standard value 0.5.

Note that these phase diagrams are very similar to Fig. 4 in
\cite{Zhang:2009zk}, which was obtained in a nonlocal two-flavor
NJL model. Actually, these two models almost reproduce the same
constituent quark ($u$ and $d$) masses and have similar scale
parameters. It reflects that the strange quark plays only a minor
role for the critical properties of the chiral phase transition.
Figure \ref{fig:pd-3flavor} shows that the KMT interaction does
not alter the possible multiple critical-point structures for the
chiral phase transition.

It is noteworthy that the vector interaction acts so as to suppress the
chromomagnetic instability of the gapless phases.
For simplicity, the unstable regions with chromomagnetic
instability are not plotted in Fig.\ref{fig:pd-3flavor}. The
calculation of the Meissner mass squared in the 2CSC phase for the
two-plus-one-flavor case is straightforward but complicated.
Including the $s$ quark should have little effect on the value of
the Meissner masses calculated according to the formula for the
two-flavor case \cite{Huang:2004bg} since the $s$ quark does not
take part in in the Cooper pairing. We can expect that the critical
points E, F,  and G should be free from the chromomagnetic
instability,  as in Fig. 4 for the two-flavor case
\cite{Zhang:2009zk}, since the large strange quark mass may give a
positive contribution rather than a negative one to the Meissner
masses squared. As for critical point H, it might be located in the
unstable region but could be safe from the instability because the
relatively large $G_V/G_S$ may shift the unstable region to
lower $T$ and higher $\mu$.

Usually,  it is argued that the instantons should be screened at
large chemical potential and temperature. Therefore, compared to
it's vacuum value, the coupling constant $K$ is expected to be
reduced around the chiral boundary. For smaller $K$, the flavor
mixing effect is suppressed and the mass mismatch between the $s$
quark and the $u(d)$ quark becomes larger. Accordingly, the
influence of the $s$ quark on the chiral restoration is weakened,
and the situation approaches the two-flavor case. On the other
hand, with decreasing $K$, the $u(d)$ quark mass also decreases
since the contribution from the $s$ quark mass is reduced. This
means that the first-order chiral restoration will be weakened
when $K$ is decreased. Correspondingly, the COE region should be
more easily formed with the combined influence of the vector
interaction and the neutral CSC phase, which favors the multiple
critical-point structures or crossover for chiral restoration at
low temperature.

Of course, the produced $u(d)$ quark vacuum constituent masses
with different model parameters of the two-plus-one-flavor NJL
model may range from 300-400 MeV, which are all phenomenologically
acceptable just as in the two-flavor case. Then, one can expect
that all the critical-point structures found in the two-flavor
case (the Sec. II of \cite{Zhang:2009zk} ) should also appear in
the two-plus-one-flavor case, even when considering the axial
anomaly interaction term.

%%%%%%%%%%   CONCLUSIONS AND OUTLOOK   %%%%%%%%%%

\section{CONCLUSIONS AND OUTLOOK}

The vector-vector interaction and electric-charge neutrality in
$\beta$ equilibrium have important effects on the chiral phase
transition. In the presence of color superconductivity(CSC), as
demonstrated in \cite{Zhang:2009zk} within the NJL model,  the QCD
phase diagram may have a multiple chiral critical-point structure.
%as shown in \cite{Zhang:2009zk},
%When the CSC phase transition is taken into consideration
%under the electric charge neutrality constraint,
The emergence of such a multiple critical-point structure is made
robust with the inclusion  of the repulsive vector coupling $G_V$
between quarks, and it becomes that there always exists a
parameter window in the NJL models which favors the appearance of
a multiple chiral critical-point structure for a wide range of the
vacuum quark mass, i.e., from 300 MeV to 400 MeV
\cite{Zhang:2009zk}. More precisely, besides the two- and
three-critical-points structures found in
\cite{KitazawaVector,Zhang:2008wx} and  \cite{Zhang:2008wx},
respectively, the QCD phase diagram can have even {\em four}
critical points; such a multiple critical-point structure is
caused by the combined effect of the positive $\mu_e$ and $G_V$.
Because the dynamical strange quark mass is still relatively large
near the boundary of the chiral transition, the multiple
critical-point structures found in the two-flavor case also appear
in the two-plus-one-flavor case. For the intermediate diquark
coupling,  the number of critical points changes  as $1\,
\rightarrow \, 2\,\rightarrow\, 4\,\rightarrow\, 2\,
\rightarrow\,0$ with an increasing vector coupling in the
two-plus-one-flavor NJL model. In general, one can expect that
different model parameters may possibly give other order of the
number of the critical points as the vector coupling is increased.

Although our analysis is based on a low-energy effective  model
which inherently has, more or less,  a parameter dependence, we
have seen that the physical mechanism to realize the multiple
critical-point structure is solely dependent on the basic
ingredients of the effective quark dynamics and thermodynamics.
Therefore, we believe that the results obtained in the present
work should be taken seriously and examined in other effective
models of QCD,  or hopefully lattice QCD simulations.

Our result also has a meaningful implication for the study of
phase transitions in condensed matter physics. That means some
external constraints enforced on the system can lead to the
formation or expansion of the coexisting phase,  and the
competition between two order parameters can give rise to multiple
critical points.

Last but not least, in \cite{Zhang:2009zk} we have shown for the
first time that the repulsive vector interaction  does suppress
the chromomagnetic instability related to the asymmetric
homogeneous 2CSC phase. Even such a conclusion is based on a
nonlocal two-flavor NJL model, it should also be true for the
two-plus-one-flavor case because of the relatively large mass of s
quark. With increasing vector interaction, the unstable region
associated with chromomagnetic instability shrinks towards lower
temperatures and higher chemical potentials. This means that the
vector interaction can partially or even totally resolve  the
chromomagnetic instability problem.

To cure the chromomagnetic instability, inhomogeneous
asymmetric color superconductivity phases such as the LOFF phase
and the gluonic phase were proposed in the literature. For the
inhomogeneous phase, beside the condensate
$\langle\overline{\psi}{\gamma_0}\psi\rangle$, there is no reason
to rule out the appearance of another new condensate ,
$\langle\overline{\psi}\vec{\gamma}\psi\rangle$,  when considering
the vector interaction. The effect of both the timelike vector
condensate and the spacelike condensate on the asymmetric
inhomogeneous  CSC phase will be reported in our future work.

\bigskip

\acknowledgements

One of the authors ( Z.~Z. ) is grateful for the support from the
Grants-in-Aid provided by Japan Society for the Promotion of
Science (JSPS). This work was partially supported by a
Grant-in-Aid for Scientific Research by the Ministry of Education,
Culture, Sports, Science and Technology (MEXT) of Japan (No.
20540265 and No. 19$\cdot$07797),
 by Yukawa International Program for Quark-Hadron Sciences, and by the
Grant-in-Aid for the global COE program `` The Next Generation of
Physics, Spun from Universality and Emergence '' from MEXT.

\end{document}